# Polarization-Tunable Perovskite Light-Emitting Metatransistor

Maciej Klein,[1,2,†] Yutao Wang,[1,2,†] Jingyi Tian,[1,2] Son Tung Ha,[3] Ramón Paniagua-Domínguez,[3] Arseniy I. Kuznetsov,[3] Giorgio Adamo[1,2] and Cesare Soci[1,2*]

[1] Centre for Disruptive Photonic Technologies, TPI, Nanyang Technological University, 21 Nanyang Link, Singapore 637371

[2] Division of Physics and Applied Physics, School of Physical and Mathematical Sciences, Nanyang Technological University, 21 Nanyang Link, Singapore 637371

[3] Institute of Materials Research and Engineering, Agency for Science Technology and Research (A*STAR), 2 Fusionopolis Way, Singapore 138634

*Correspondence to: csoci@ntu.edu.sg

† These authors contributed equally

**Emerging immersive visual communication technologies require light sources with complex functionality for dynamic control of polarization, directivity, wavefront, spectrum, and intensity of light. Currently, this is mostly achieved by free space bulk optic elements, limiting the adoption of these technologies. Flat optics based on artificially structured metasurfaces that operate at the sub-wavelength scale are a viable solution, however their integration into electrically driven devices remains challenging. Here we demonstrate a radically new approach of monolithic integration of a dielectric metasurface into a perovskite light-emitting transistor. We show that nanogratings directly structured on top of the transistor channel yield an 8-fold increase of electroluminescence intensity and dynamic tunability of polarization. This new light-emitting *metatransistor* device concept opens unlimited opportunities for light management strategies based on metasurface design and integration.**

Light-emitting devices have become an integral part of our daily lives, where they cater to a wide range of needs, from ambient illumination in homes and workplaces to displays in televisions, mobile phones, and computer screens. With the advent of immersive visual technologies, such as virtual/augmented reality (VR/AR) and 3D holographic displays,[1] there is a rapidly growing demand for light-emitting devices that incorporate complex functionalities to dynamically manipulate intensity, wavelength, polarization, phase, and wavefront of the emitted light. Dielectric metasurfaces, two-dimensional planar arrangements of low-loss nanoscale optical elements, are prime candidates for this endeavor: by tight confinement and manipulation of the electromagnetic fields at the subwavelength scale, they allow exerting passive and active control over the amplitude, polarization and phase of reflected, transmitted or emitted light, beyond what is possible by conventional optical elements.[2–4]

Thanks to their compatibility with a variety of nanofabrication techniques and high refractive indices, organic-inorganic halide perovskites are an ideal platform for dielectric metasurfaces. Perovskite metamaterials and metasurfaces[5–7] for structural coloring,[8] light-emission[9–11] and chirality[12] control, dynamic color displays,[13,14] and lasers[15,16] operating across the visible and infrared spectral range have been demonstrated recently. In addition, the unique optoelectronic properties of halide perovskites, namely high luminescence efficiency, wide-range bandgap tunability and exceptional color purity, long carrier lifetime and diffusion length, and high tolerance to defects,[17] make them an extremely flexible materials platform for electrically driven light-emitting devices, with perovskite light-emitting diodes (PeLED) already reaching external quantum efficiencies in excess of 20%.[18]

While several groups reported direct patterning of functional metasurfaces in the active materials of optically pumped devices,[19–22] light management strategies based on hybridization of functional metasurfaces in electrically driven light-emitting devices, such as light-emitting diodes (LEDs) and light-emitting transistors (LETs),[23] are just in their infancy. Inorganic LEDs with integrated metasurfaces have shown highly polarized electroluminescence (EL) emission[24], while light extraction and color purity improvement, as well as wavefront and polarization control, were attained in organic light-emitting diodes.[25–27] The incorporation of plasmonic nanostructures on the glass substrate of organic light-emitting transistors led to directional electroluminescence.[28] Similarly, directional polarized light emission was also achieved in perovskite light-emitting diodes (PeLEDs) with patterned substrates.[29,30] So far, all demonstrations of hybridized metasurfaces relied on nanostructured layers residing outside the active region of the light-emitting device. This is due to the poor manufacturability of the active materials as well as to the limitations imposed by the vertical device architecture of conventional LEDs, where additional interlayers may totally disrupt charge carrier transport and radiative recombination pathways within the device.

Here we present a monolithic perovskite light-emitting metasurface transistor, or metatransistor (PeLEMT), where the metasurface is directly patterned into its channel (**Fig. 1**). This is made possible by the lateral design configuration of the PeLET, which borrows its architecture from conventional field-effect transistors (FETs) and emits light from the open surface between top electrodes.[31–33] Metasurfaces consisting of arrays of nanogratings, with resonant response that can be tuned across the electroluminescence spectrum of the LET, lead to almost one order of magnitude spectral enhancement of the electroluminescence. In addition, exploiting polarization properties of the nanograting and spatial control of the recombination zone by electrical bias, we demonstrate dynamic tunability of the emission polarization from a pixelated metasurface.

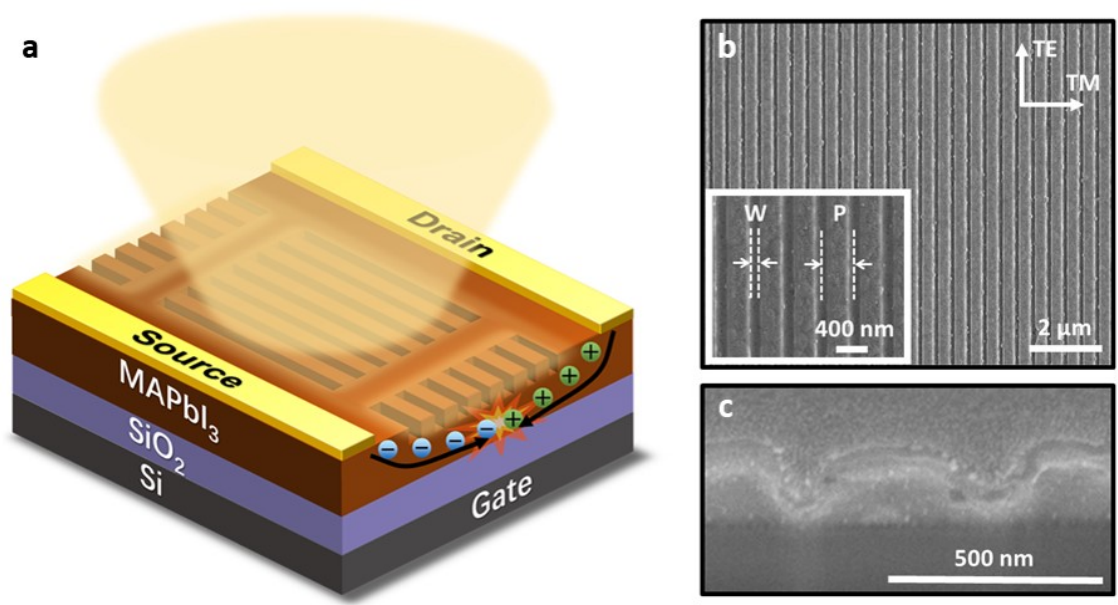

**Figure 1. Perovskite light-emitting metatransistor.** (a) Schematic illustration of a perovskite light-emitting metatransistor incorporating a monolithic metasurface within its surface active area. (b) Top-view and (c) cross-sectional SEM images of a MAPbI$_3$ perovskite film inscribed with a nanograting metasurface. The inset shows an enlarged view of the fabricated structure with its geometrical parameters: groove width (W) and nanograting period (p).

**Transport characteristics of nanopatterned transistors**

Devices used in this study are based on an optimized bottom-gate, top-contact light-emitting transistor configuration, and methylammonium lead iodide (CH$_3$NH$_3$PbI$_3$) MAPbI$_3$ perovskite as the light-emitting dielectric medium.[32] ~200 $nm$ thick MAPbI$_3$ films are spun-cast on $p$-doped Si substrates covered by 500 $nm$ thick thermally grown silicon oxide layer (acting as bottom gate and gate insulator, respectively) and patterned with nanograting metasurfaces by Focused Ion Beam (FIB) milling. The monolithic metadevice architecture and working principle are shown in **Fig. 1a.** We manufactured nine ~$30 \times 30\ \mu m$ nanograting metasurfaces within the LET active area, with fixed groove width $W \approx 100$ nm, varying period $p=[400 - 480]\ nm$, in steps of $10\ nm$, and depth $d\sim 150\ nm$ (**Fig. 1b**). We ensured that the FIB milling left a ~$50\ nm$ residual layer close to the gate electrode (**Fig. 1c**), which is essential for the inversion layer (the so-called channel) at $MAPbI_3/SiO_2$ interface to be formed and permit efficient, gate field modulated, charge carrier transport between source and drain electrodes.[34] A small region of the perovskite film near the electrodes is left unpatterned to limit disruption of charge injection at the metal/semiconductor interface, avoid the formation of short circuits or high conductance pathways within the device active region, and reduce electrode-induced photon losses. To reduce the field screening effects arising from ionic transport in $MAPbI_3$ that are known to limit carrier mobility and induce large electrical hysteresis at room temperature,[31] all our electrical and optical measurements were carried out at $77\ K$.

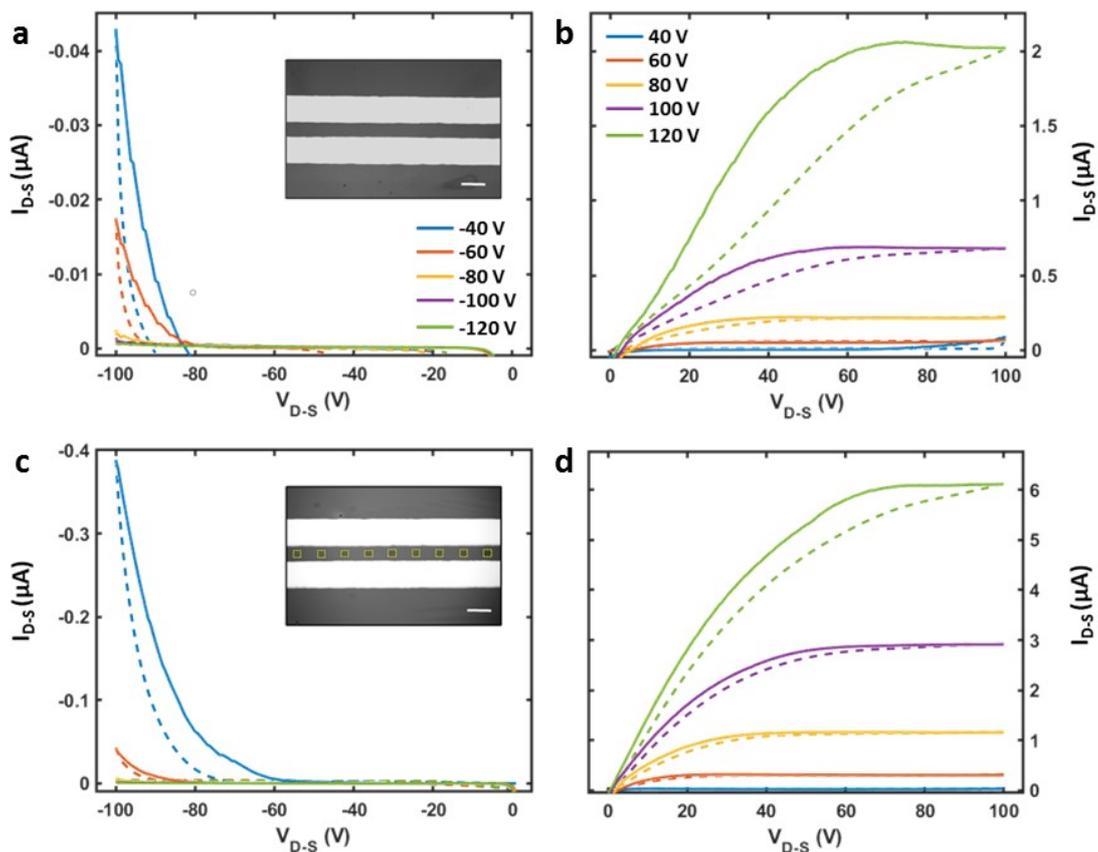

**Figure 2. Electrical transport characteristics of light-emitting transistors.** (a,c) *p*-type and (b,d) *n*-type output characteristics of $MAPbI_3$ LETs obtained before (a,b) and after (c,d) metasurface fabrication. The LET characteristics were measured at $77\ K$ under constant gate-source bias of $V_{G-S}=40$ to $120\ V$ for *n*-type transport and $V_{G-S}=-40$ to $-120\ V$ for *p*-type transport. Solid lines were obtained in the forward direction while dashed lines in the reverse

direction of the voltage sweep. Insets: top-view optical microscope images of (a) the unstructured LET and (c) the LET structured with an array of square metasurfaces (highlighted by yellow dashed lines).

Electrical output and transfer characteristics in **Fig. 2** and **Fig. S1** in Supplementary Information reveal that FIB nanostructuring does not have detrimental effects on the transistor performance. The current-voltage ($I_{D-S}$–$V_{D-S}$) output curves of the transistor before (**Fig. 2a, b**) and after (**Fig. 2c, d**) inscription of the metasurfaces show the typical ambipolar character of charge transport in MAPbI3, with electrons dominating conduction. Electrical hysteresis of the current between forward and reverse bias sweeps, typical of halide perovskites with ionic character[31,35], is visible. The field-effect electron ($\mu_e$) and hole ($\mu_h$) mobilities extracted for the metatransistor at the drain-source voltage of $V_{D-S} = 80\ V$ ($\mu_e = 9.5 \times 10^{-2}\ cm^2V^{-1}s^{-1}$ and $\mu_h = 6.4 \times 10^{-5}\ cm^2V^{-1}s^{-1}$) are comparable to those of unstructured devices ($\mu_e = 7.5 \times 10^{-2}\ cm^2V^{-1}s^{-1}$ and $\mu_h = 3.8 \times 10^{-4}\ cm^2V^{-1}s^{-1}$). Output curves show strong gate modulation with a distinguishable *n*-type linear and saturation work regime. Significantly higher drain-source currents ($I_{D-S}$) with lower electrical hysteresis were obtained after patterning, which may indicate an improvement of carrier density induced by Ga ions implantation upon MAPbI3 nanofabrication.

**Polarization-enhanced metatransistor electroluminescence**

Along with improved transport characteristics, the perovskite light-emitting metatransistor showed uniform and stable, channel-wide electroluminescence. The EL intensity is clearly enhanced in the emitting regions overlapping with the metasurface array, as seen in the optical image in **Fig. 3a**. To study the dependence of EL enhancement and its polarization on the geometrical parameters of the nanogratings, we collected the light emitted by each metasurface through a 10x microscope objective and a polarizer. The optical modes supported by the metasurfaces are revealed by detecting two orthogonal polarizations, parallel (transverse electric, TE) and perpendicular (transverse magnetic, TM) to the nanogratings. The TE and TM electroluminescence spectra from metasurfaces of period increasing from 400 to 480 $nm$ are shown in **Fig. 3b** and **Fig. 3c**, overlaid to the EL of the unpatterned film. The first clear manifestation of the interaction with the metasurfaces is the EL intensity enhancement induced for both TE and TM polarizations, with a distinct spectral dependence on the nanograting period. The resonant modes appear as peaks in both the TE and TM electroluminescence spectra that redshift, more or less rapidly, when the nanograting period increases from 400 to 480 $nm$ (the modes are tracked with dashed lines in **Fig. 3b** and **Fig. 3c**, which serve as guides to the eye). The same modes manifest, for both TE and TM, in photoluminescence, when the MAPbI$_3$ metasurfaces are optically excited by a 405 nm laser (**Fig. S2** in Supplementary Information). The excellent correspondence of the metasurface modes in photo- and electroluminescence spectra indicates that the processes underlying light-matter interaction in the light-emitting nanogratings upon charge injection or optical excitation are the same, that is the metasurface is optically coupled to the active region of the light-emitting transistor and does not merely act as an optical filter. The EL enhancement produced by the metasurfaces can be quantified by the ratio between the intensity of light emitted by the arrays and by the unstructured film, under the same electrical bias conditions. The colormap plot in **Fig. 3d** shows the EL enhancement as function of both emission wavelength and nanogratings period: two spectral branches corresponding to dominant metasurface modes can be clearly identified as a function of the grating period. An 8-fold enhancement of the EL intensity is reached at $\lambda = 792\ nm$ for a nanograting with period $p = 480\ nm$. Based on the ratio between polarized emission intensities, $I_{TE}/I_{TM}$, plotted in **Fig. 3e**, the origin of the two branches can be attributed to the dominant TE modes of the metasurfaces. This is consistent with the photoluminescence characteristics of the nanogratings, where the TE modes are those responsible for the observed enhancement (**Fig. S2c, d** in Supplementary Information).

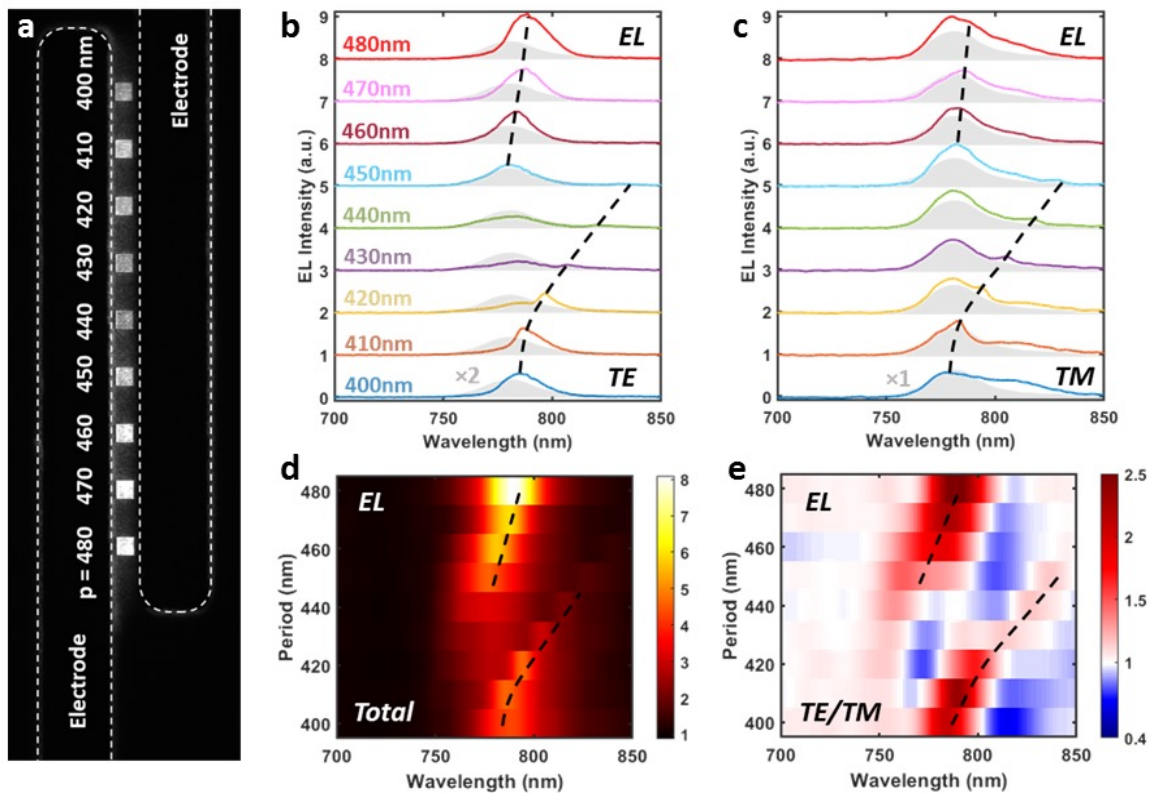

**Figure 3. Electroluminescence of the perovskite light-emitting metatransistor.** (a) Optical image of the EL emitted from the PeLEMT channel patterned with nanograting metasurfaces

of periods increasing from 400 to 480 $nm$. Normalized (b) TE and (c) TM EL spectra. The shaded spectra are the reference EL from the unpatterned perovskite film. Colormaps of (d) the total EL intensity enhancement and (e) the TE/TM intensity ratio as a function of emission wavelength and nanograting period. The dashed lines are guides to the eye for the optical modes. The transistor was biased at $V_{D-S} = 100\ V$ and $V_{G-S} = \pm 60\ V$, with $100\ kHz$ modulation frequency. Measurements were carried out at $77\ K$.

Both EL and PL enhancements can be attributed to a combination of better outcoupling induced by the nanograting and the Purcell effect determined by the confinement of optical modes within the perovskite nanogratings.[8,11,36] A combination of full-wave numerical simulations and angle-resolved back focal plane spectral measurements, in reflection and EL emission, helped to unveil the nature of the resonant modes.

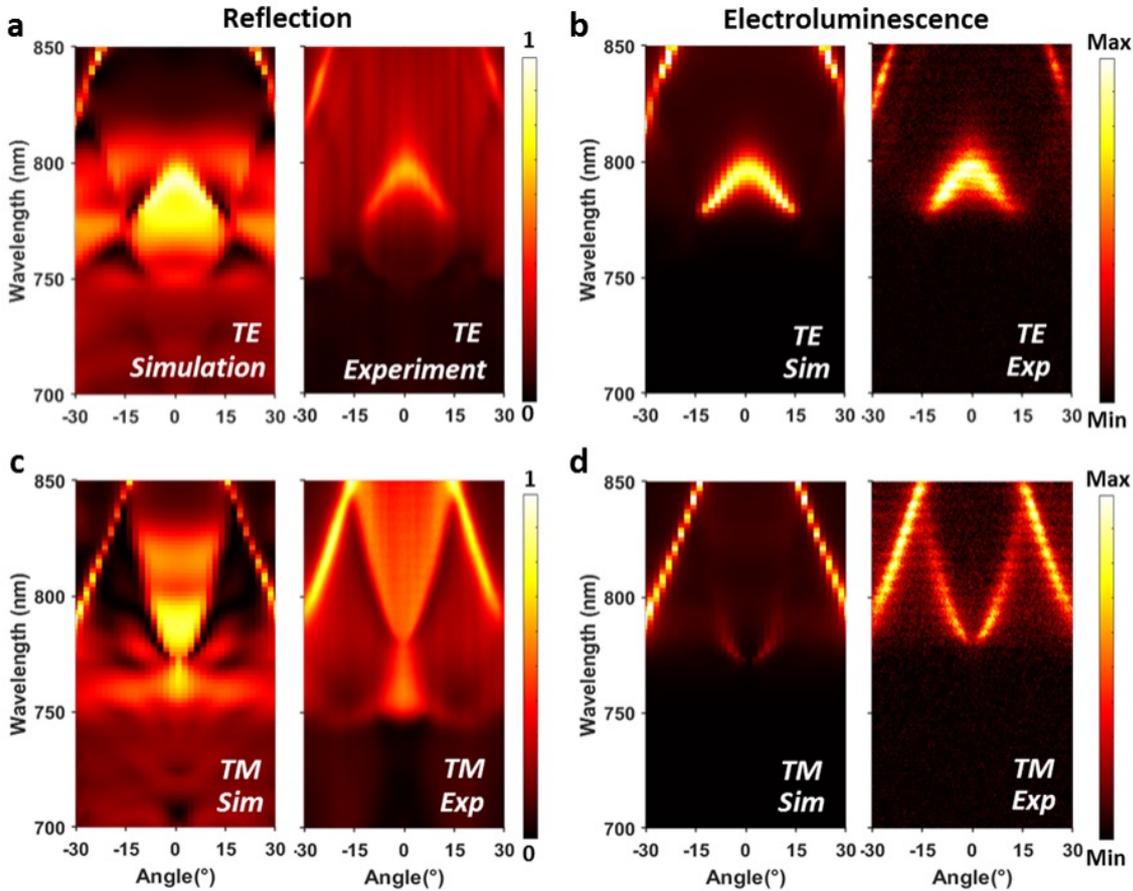

**Figure 4. Optical modes in the metatransistor unveiled by angle-resolved spectral maps.** (a) Reflection and (b) EL angle-resolved TE spectral maps for the $p = 480\ nm$ nanograting metasurface, revealing the optical mode responsible for the $\lambda = 790\ nm$ resonant emission, in both numerical simulation and experiments. (c) Reflection and (d) EL angle-resolved TM spectral maps for the $p = 480\ nm$ nanograting metasurface, revealing the quasi-BIC responsible for the $\lambda = 780\ nm$ resonant emission, in both numerical simulation and experiments. The transistor was biased at $V_{D-S} = 100\ V$ and $V_{G-S} = \pm 60\ V$, with $100\ kHz$ modulation frequency. Measurements were carried out at $77\ K$.

The $\lambda = 790\ nm$ TE resonance is responsible for the highest enhancement given by the $p = 480\ nm$ nanograting, that yields reflection and far-field EL emission mostly along the normal direction (**Figs. 4a** and **4b**). The $\lambda = 780\ nm$ TM resonance mode, supported by same structure, is a quasi-bound state in the continuum (quasi-BIC)[16,37,38] known for its high radiative quality factor. The dispersion of quasi-BICs is characterized by high reflection and bright emission surrounding an area of vanishing intensity around the normal direction, a feature well captured in the numerical simulations and experimental maps of **Fig. 4c, d**. The nearfield distribution of the electric and magnetic fields of the TE and TM modes (**Figs. S3** in Supplementary Information) reveal the strong field confinement within the perovskite metasurface, that accounts for the observed enhancement. The nature of the high brightness TE and TM resonances remains the same while reducing the nanogratings period from $p = 480\ nm$ to $p = 440\ nm$ and correspond to the upper branches identified in **Fig 3**, after which they blue-shift and disappear due to the high MAPbI$_3$ absorption at $\lambda < 750\ nm$. Concurrently, modes of different nature for both TE (quasi-BIC) and TM polarizations become dominant in nanogratings with $p \leq 440\ nm$ (**Fig. S4** in Supplementary Information), matching the long wavelength branches identified in **Fig 3**. Photoluminescence measured in back focal plane imaging shows spatial patterns consistent with those recorded for reflection and electroluminescence (**Fig. S5** in Supplementary Information).

**Dynamic pixel-polarization switching**

A unique advantage of the 3-terminal device architecture of LETs over 2-terminal light-emitting devices such as LEDs is the possibility to control the spatial location of the light emission zone. By operating the device with AC bias and varying $V_{D-S}$, it is possible to arbitrarily position the recombination zone between the source and the drain terminals.[32] This feature opens to the possibility to dynamically address different patterns in a spatially pixelated metasurface array. As proof of principle, here we demonstrate dynamic switching of the light polarization emitted from two nanogratings placed at different locations of the metatransistor channel. The nanogratings have the same period ($p = 420\ nm$) and orthogonal orientation, giving the highest TE/TM contrast of 2.5 at $\lambda = 790\ nm$. Optical images of the EL from the patterned device clearly show emission from either one of the two metasurface pixels when $V_{D-S}$ is swept between $80\ V$ and $10\ V$ and the recombination zone moves across the device channel (**Fig. 5a**). Correspondingly, the pixelated metatransistor channel yields distinct polarized emission that can be controlled by the amplitude of the electrical bias. Polarized EL collected by a 10x microscope from the centre of the device channel (**Fig. S6** in Supplementary

Information) shows modulation of the peak intensity around $\lambda = 790\ nm$ while changing the drain-source bias from $V_{D-S} = 80\ V$ to $V_{D-S} = 10\ V$ (**Fig. 5b**). The intensity of polarized EL emitted from each of the two orthogonal metasurfaces is characterized by collecting the total emission through varying linear polarizer orientations (**Fig. 5c**), showing that the ratio of EL intensities emitted by the two pixels can be varied continuously. The degree of polarization of the electroluminescence (**Fig. 5d**), $DOP = (I_0 - I_{90})/(I_0 + I_{90})$, where $I_0$ and $I_{90}$ are EL intensities at 0 and 90° polarization, reaches the maximum value of $DOP \approx \pm 0.35$ around the EL emission peak ($\lambda = [770, 810]\ nm$), and maximum value of $DOP \approx \mp 0.35$ on the long wavelength EL emission tail ($\lambda = [830, 880]\ nm$).

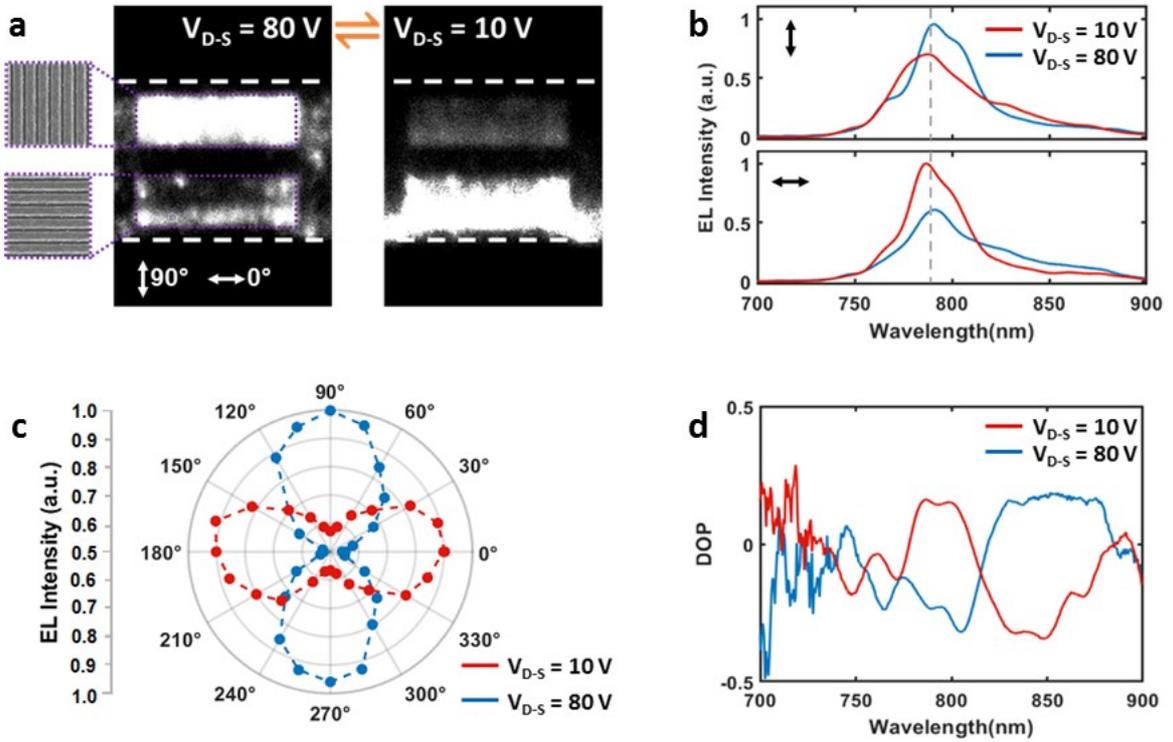

**Figure 5. Dynamic tunability of the pixel electroluminescence polarization.** (a) Optical microscope images of the EL emitted by a pixelated PeLEMT channel (within the white dashed lines, indicating the edge of the top electrodes) for drain-source voltage of $V_{D-S} = 80\ V$ (left) and $V_{D-S} = 10\ V$ (right). The pixels comprise of two nanogratings oriented in opposite directions (SEM images of the vertical and the horizontal nanogratings are shown in the insets on the left). Varying $V_{D-S}$ allows to switch between the two pixels with opposite light-emitting polarization. (b) EL emission spectra obtained under $V_{D-S} = 80\ V$ (blue curves) and $V_{D-S} = 10\ V$ (red curves) biasing conditions and measured in the vertical (top graph) and horizontal (bottom graph) polarization. (c) Polar plot of the polarized EL intensity at $\lambda = 790\ nm$ obtained for both $V_{D-S}$ biasing conditions. (d) EL degree of polarization (DOP) spectrum obtained for both $V_{D-S}$ biasing conditions. In all measurements the emitted light was collected from the two nanogratings simultaneously (see **Fig. S6** in Supplementary Information). The transistor was biased at $V_{G-S} = \pm 30\ V$, with $100\ kHz$ modulation frequency. Measurements were carried out at $77\ K$.

**Conclusions**

We have proposed and experimentally realized the monolithic integration of functional dielectric metasurfaces within the active channel of an electrically driven light-emitting transistor. The LET offers a number of advantages linked to its open surface architecture, that allows direct nanopatterning of the active area, and spatially tunable recombination zone, that allows electrical addressing of individual pixels of a metasurface array within the transistor channel by controlling the amplitude rather than the timing of the electrical bias pulses. Furthermore, charge carrier densities obtainable in LETs is several orders of magnitude greater than in conventional LEDs,[39] which may lead to high brightness devices[40] and possibly electrically-driven lasers.[41] Direct nanopatterning of light-emitting metatransistors demonstrated here using FIB lithography could be easily up-scaled by nanoimprint lithography techniques that are routinely employed to realize perovskite metasurfaces.[7] The light-emitting metatransistor device opens unlimited opportunities for integration of functional metasurfaces into electrically driven light-emitting devices. Here we demonstrated higher than 8-fold spectral enhancement and dynamic polarization switching of the electroluminescence with DOP of about 35% in pixelated devices. Furthermore, the simplicity of design and nanomanufacturing of perovskite dielectric metasurfaces will allow straightforward extension of the metatransistor concept to a plethora of functionalities for emerging visual communication technologies.

**Methods**

*PeLET fabrication:* PeLETs were fabricated in a bottom-gate and top-contact configuration on heavily *p*-doped Si substrates with thermally grown SiO$_2$ (500 $nm$) layer as the gate insulator (capacitance of $C_i = 6.9\ nF cm^{-2}$). MAPbI$_3$ thin films were fabricated from 1.2 M precursor solution of CH$_3$NH$_3$I (Dyesol) and PbI$_2$ (99.99%, TCI) (molar ratio 1:1) in anhydrous dimethylformamide (DMF, Sigma-Aldrich). As prepared solution was magnetically stirred overnight at room temperature in N$_2$ filled glovebox, then filtered by a polyvinylidene fluoride (PVDF) syringe filter (0.45 μm) and left on the hot plate at 373 K for one hour before spin-coating. Prior to perovskite deposition, substrates were cleaned with ultrasonication for 15 min in acetone, isopropanol and deionized water. Subsequently, substrates were dried with the flow of nitrogen followed by oxygen plasma cleaning treatment. The perovskite precursor solution was spin-coated onto the quartz substrates with a speed of 4900 rpm for 30 s using antisolvent engineering method. Toluene was drop-casted on the substrates 4-5 s after the start of the spin-coating program. The resulting films were finally annealed at 373 $K$ for 15 min. Afterwards, $100\ nm$ thick Au electrodes were thermally evaporated through a shadow mask in a high vacuum ($\sim 10^{-6}$ mbar), with a deposition rate of $0.5\ Ås^{-1}$. To avoid thermal decomposition of the perovskite films, samples were placed on a water-cooled substrate holder kept at $291\ K$ during electrode deposition. The resulting PeLET channel length ($L$) and width ($W$) were $60\ \mu m$ and $1\ mm$, respectively.

*Metamaterial fabrication:* To complete the PeLEMT devices, arrays of grating metasurfaces were patterned on a perovskite film, between top Au electrodes, by a focused ion beam (Helios 600 NanoLab, FEI). The lateral dimension of each fabricated structure was $30 \times 30\ \mu m$.

*Electrical and optical characterization:* Electrical and optical characterization was carried out at $77\ K$, in the dark and under vacuum ($\sim 10^{-3}\ mbar$) using a temperature-controlled probe stage (HFS600E-PB4/PB2, Linkam). DC-driven characteristics were acquired with a 2-channel precision source/measure unit (B2902A, Agilent). Charge-carrier mobilities were extracted from the forward sweeping of transfer characteristics obtained at $V_{D-S} = \pm 80\ V$, using the conventional equations for metal-oxide semiconductor (MOS) transistors in the saturation regime: $\mu_{sat} = \frac{2L}{WC_i}\left(\frac{\partial \sqrt{I_{D-S}}}{\partial V_{G-S}}\right)^2$.

Photoluminescence spectra were measured using a micro-PL setup based on an optical microscope (Eclipse LV100, Nikon) with LU plan fluor $\times 10$ and $\times 50$ objectives. A picosecond laser diode $\lambda = 405\ nm$ (P-C-405B, Picoquant) with $40\ MHz$ repetition rate was used as the excitation source while the PL emission was detected by a fiber-coupled spectrometer (AvaSpec ULS-RS-TEC, Avantes).

Electroluminescence measurements were performed on the same optical setup, under AC-driven mode by applying constant drain-source bias ($V_{D-S} = 100\ V$) and square wave bias on the gate electrode ($V_{G-S} = \pm 60\ V$ and $f = 100\ kHz$ modulation frequency), using an arbitrary waveform generator (3390, Keithley) coupled with a high-voltage amplifier (WMA-300, Falco Systems). Optical images and videos were taken and acquired by a cooled sCMOS scientific camera (PCO.edge 3.1m) coupled to the optical microscope. Dynamic pixel-polarization switching EL measurements were carried out using an optical microscope (PSM-1000, Motic) with PLAN APO $\times 10$ objective, coupled to the same spectrometer and camera. The gate bias was $V_{G-S} = \pm 30\ V$ and the modulation frequency $f = 100\ kHz$.

*Angle-resolved measurements:* Angle-resolved reflection, photoluminescence, and electroluminescence measurements were performed with a home-built micro-spectrometer. The system consists of an inverted optical microscope (Ti-U, Nikon), a spectrograph (Andor SR-303i with 150 lines/mm grating), and an electron-multiplying charged-coupled detector (EMCCD, Andor Newton 971). A lens system along the light path between the microscope and the spectrograph was used to project the back focal plane of the collection objective (Nikon ×50, with numerical aperture NA=0.55) onto the slit of the spectrograph. This configuration allows spectral measurement with angular information corresponding to the NA of the objective. A linear polarizer placed on the optical path of the lens system was used to control the collection polarization. A halogen lamp was used as an excitation source in reflection measurements while in PL measurements samples were excited by the 488 $nm$ line of a cw solid state laser.

*Numerical simulations:* The angle-resolved reflection spectra were computed in Comsol Multiphysics 5.4. Floquet periodic boundary conditions were used in the transverse direction, with excitation by a periodic port above the array, and a perfectly matched layer followed by a scattering boundary after the silicon substrate. The angle resolved EL spectra were calculated, by reciprocity, integrating the energy ($\sim|E|^2$) inside the perovskite grating metasurface at each angle of reflection, from -30 to 30°.

**Acknowledgments**

This research was supported by the Agency for Science, Technology and Research A*STAR-AME programmatic grant on Nanoantenna Spatial Light Modulators for Next-Gen Display Technologies (Grant no. A18A7b0058) and the Singapore Ministry of Education MOE Tier 3 (Grant no. MOE2016-T3-1-006).

**Data availability**

The authors declare that all data supporting the findings of this study are available within this article and its supplementary information and are openly available in NTU research data repository DR-NTU (Data) at https://doi.org/XXXXXX. Additional data related to this paper may be requested from the authors.

**Author contributions**

C.S., M.K. and G.A. conceived the idea. M.K. and G.A. developed the PeLEMT fabrication procedure (G.A. carried out FIB fabrications). M.K. carried out electrical and optical measurements together with Y.W.. Y.W. performed the numerical simulations under the supervision of G.A. and J.T.. S.T.H. performed the angle-resolved measurements with support of M.K. and Y.W.. M.K., G.A. and C.S. drafted the manuscript. All the authors discussed the results and contributed to finalizing the manuscript. C.S. supervised the work.

Supplementary Information for

# Polarization-Tunable Perovskite Light-Emitting Metatransistor

Maciej Klein,[1,2,†] Yutao Wang,[1,2,†] Jingyi Tian,[1,2] Son Tung Ha,[3] Ramón Paniagua-Domínguez,[3] Arseniy I. Kuznetsov,[3] Giorgio Adamo[1,2] and Cesare Soci[1,2*]

[1] Centre for Disruptive Photonic Technologies, TPI, Nanyang Technological University, 21 Nanyang Link, Singapore 637371

[2] Division of Physics and Applied Physics, School of Physical and Mathematical Sciences, Nanyang Technological University, 21 Nanyang Link, Singapore 637371

[3] Institute of Materials Research and Engineering, Agency for Science Technology and Research (A*STAR), 2 Fusionopolis Way, Singapore 138634

*Correspondence to: csoci@ntu.edu.sg

† These authors contributed equally

Available supplementary information:

**Figure S1.** Electrical transfer characteristics of LETs.

**Figure S2.** Photoluminescence spectra of the PeLEMT.

**Figure S3.** Simulated electric and magnetic field distributions.

**Figure S4.** Angle-resolved reflection spectra of the PeLEMT.

**Figure S5.** Angle-resolved photoluminescence of the PeLEMT.

**Figure S6.** The collection area of the EL for polarization tunability measurements.

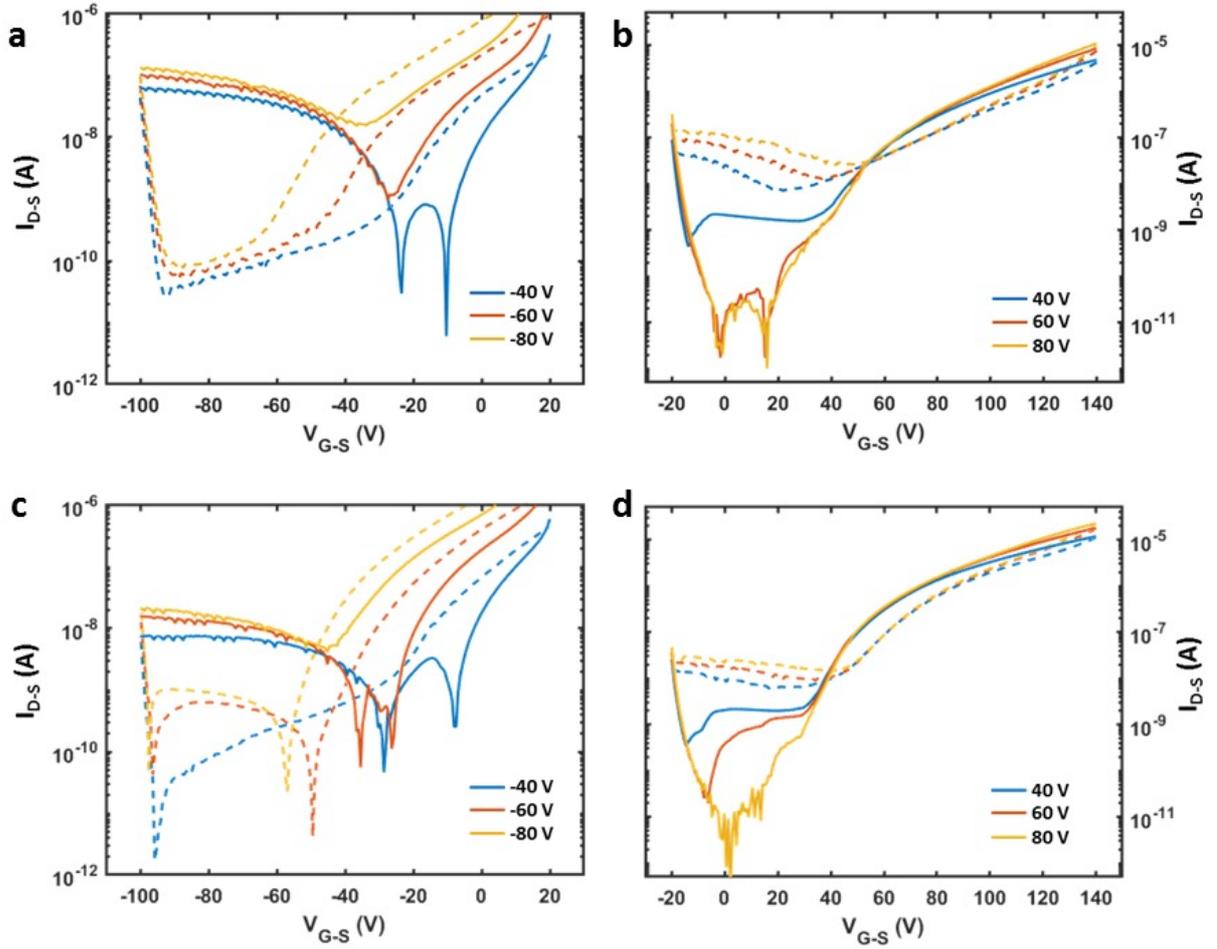

**Figure S1.** (a,c) *p*-type and (b,d) *n*-type transfer characteristics of LETs before (a,b) and after (c,d) metasurfaces fabrication. Characteristics were measured at 77 K, under the constant drain-source bias ($V_{D–S}$) from 40 to 80 $V$ for *n*-type transport ($-40$ to $-80$ $V$ for *p*-type transport) as indicated in the panels. Solid lines are obtained in the forward direction of the voltage sweep, while the dashed lines are obtained in the reverse voltage sweep direction.

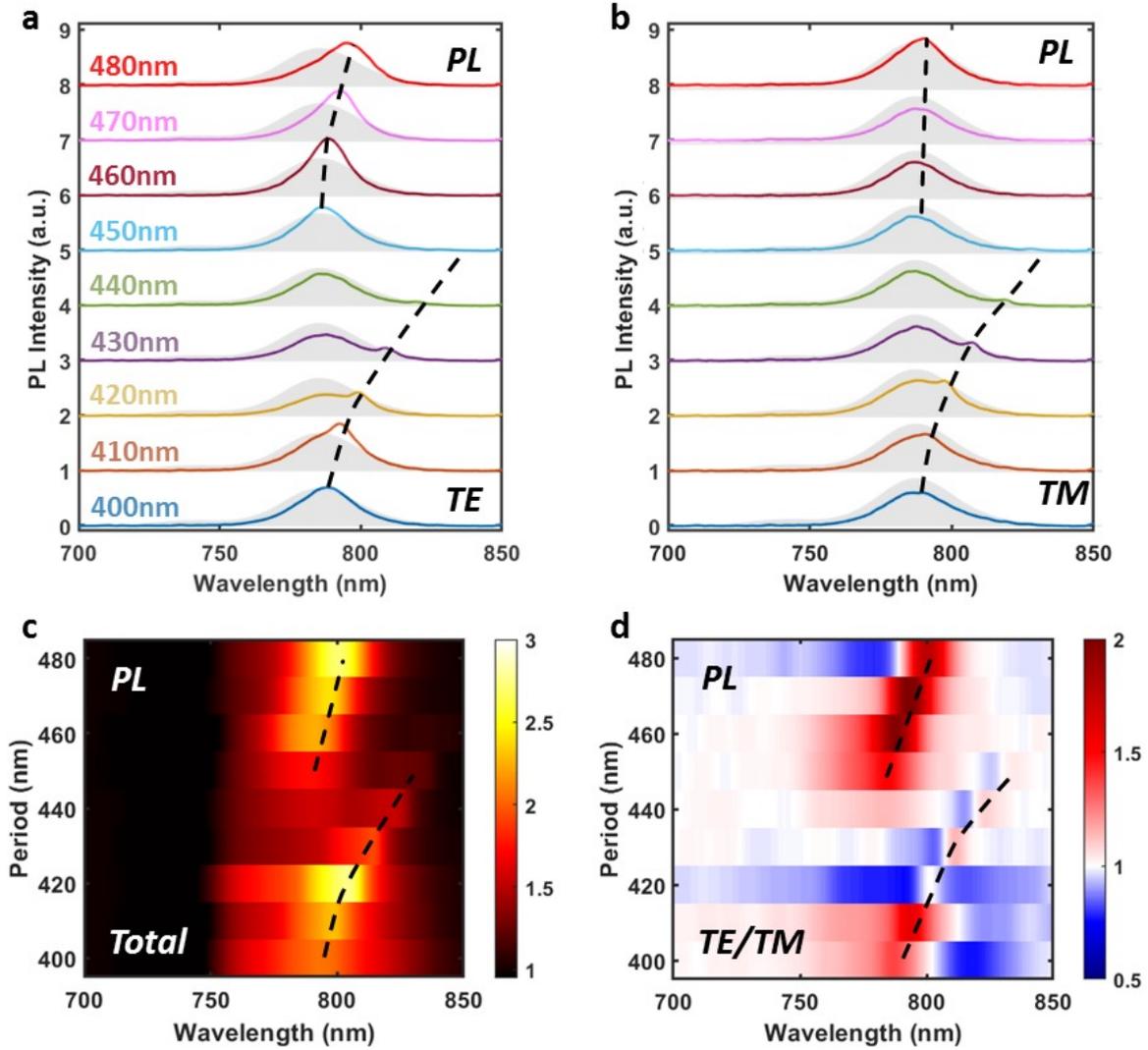

**Figure S2.** Photoluminescence spectra of (a) TE and (b) TM polarized light emitted by the metasurfaces at 77 $K$. Shaded spectra represent PL emission from the unstructured film. All spectra were normalised with regard to the maximum intensity of a given data set to conserve their relative intensity. 2D contour plot of the period-dependent (c) total PL intensity enhancement and (d) TE/TM polarized PL spectra. Dashed lines in the panels are guides to the eye for resonances shift.

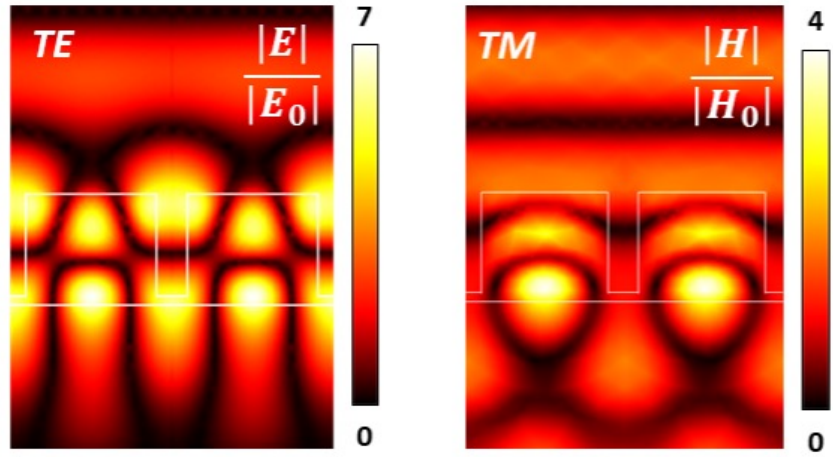

**Figure S3.** Simulated electric and magnetic field distributions in the $p = 480\ nm$ nanograting for TE- and TM-polarized incident light of $\lambda = 780\ nm$, obtained using the optical constants of MAPbI$_3$ at $77\ K$.

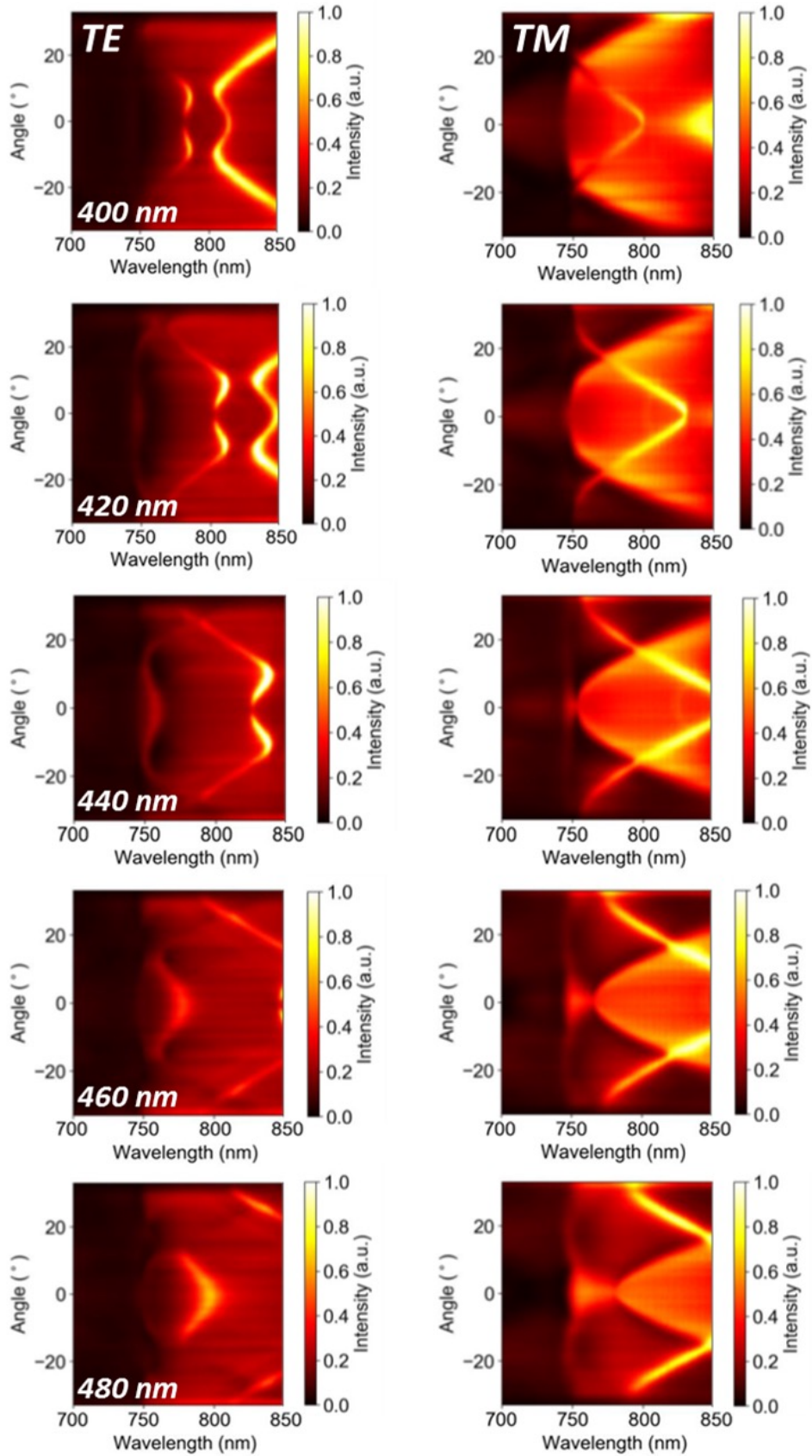

**Figure S4.** Angle-resolved reflection spectra of nanograting structures with period spanning from 400 to 480 *nm*, measured for TE (left column) and TM (right column) polarizations at 77 *K*.

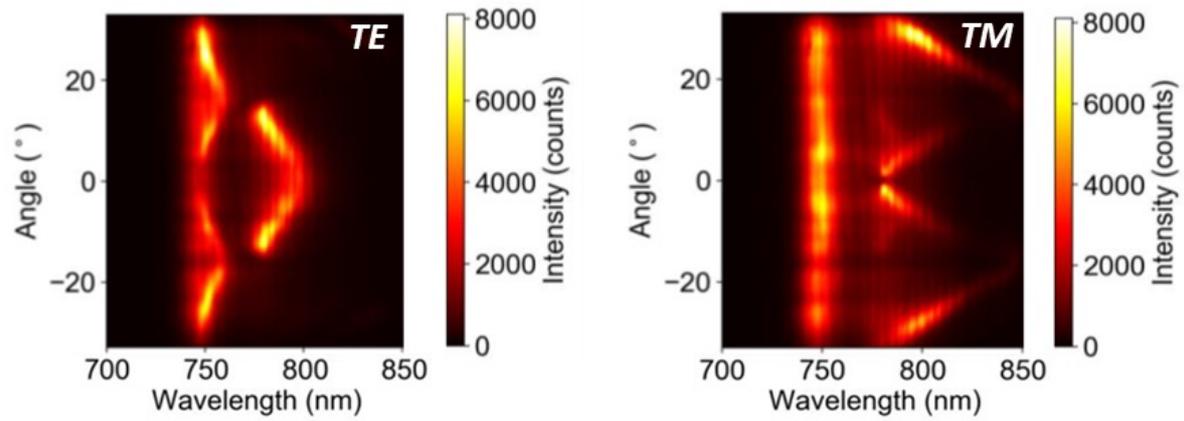

**Figure S5.** TE- and TM-polarized angle-resolved photoluminescence for nanograting structures with period $p = 480\ nm$ at $77\ K$.

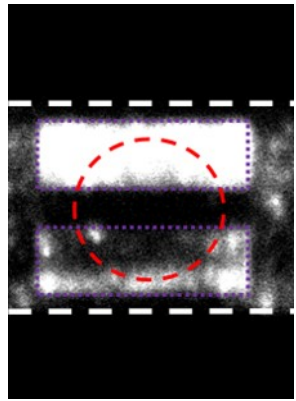

**Figure S6.** Optical microscope image of the EL emitted by the pixelated PeLEMT discussed in Fig. 5. The red circle illustrates the collection area of the EL signal that is defined by the optical fibre.